# Distinct Roles of Hydrogen in Superconducting and Ferromagnetic Phases of $CoZr_2H_x$

Yuto Watanabe,[1]* Takayoshi Katase,[2,3] Daisuke Takegami,[1,4] Yoshikazu Mizuguchi[1†]

[1]*Department of Physics, Tokyo Metropolitan University, 1-1 Minami-Osawa, Hachioji, Tokyo 192-0397, Japan*
[2] *Materials and Structures Laboratory, Institute of Integrated Research, Institute of Science Tokyo, 4259 Nagatsuta, Midori-ku, Yokohama, Kanagawa 226-8501, Japan*
[3] *MDX Research Center for Element Strategy, Institute of Integrated Research, Institute of Science Tokyo, 4259 Nagatsuta, Midori-ku, Yokohama, Kanagawa 226-8501, Japan*
[4]*Max Planck Institute for Chemical Physics of Solids, Nöthnitzer Straße 40, 01187 Dresden, Germany*

**ABSTRACT**. Hydrogenation offers a versatile route to tuning the physical properties of intermetallic compounds. In this study, we synthesized $CoZr_2H_x$ with different hydrogen contents and found that hydrogen is incorporated in two distinct concentration regimes separated by a wide composition gap: a low-concentration hydrogenated superconducting phase ($x$ = 0–0.054) and a high-concentration hydrogenated ferromagnetic phase ($x$ = 2.786). Hydrogen plays fundamentally different roles in the two concentration regimes. In the high hydrogen concentration phase, the Zr-H interactions substantially modify the metallic bands crossing the Fermi level, leading to the emergence of ferromagnetism. In contrast, in the low hydrogen concentration phase, hydrogen behaves as a nonmagnetic impurity without altering the electronic band structure. Despite the nearly identical Debye temperatures across the low-concentration series, the superconducting transition temperature ($T_c$) is progressively suppressed with increasing hydrogen content. The observed $T_c$ suppression is quantitatively described by the Abrikosov–Gor'kov pair-breaking theory, indicating that the superconducting gap of $CoZr_2$ is anisotropic or multigap rather than a fully isotropic *s*-wave symmetry.

## I. INTRODUCTION.

Hydrogen incorporation is a versatile approach to tuning the physical properties of materials through modifications to crystal and electronic structures [1,2]. For example, hydrogenated $YH_x$ shows a metal-insulator transition depending on the hydrogen concentration [3]. In contrast, the exact opposite effect occurs with the non-superconducting Pd, which upon hydrogenation becomes $PdH_x$, one of the best-known hydride superconductors. [4,5] In $PdH_x$, it is considered that superconductivity arises due to the suppression of the spin fluctuations, present in the enhanced Pauli paramagnetic Pd metal [4]. Superconductivity of $PdH_x$ emerges over $x \sim 0.7$, and its superconducting transition temperature $T_c$ reaches around 9 K for $x \sim 1$ [6]. Hydrogenation can also serve as a dopant, for instance, in iron-based superconductors [7–9]. For instance, in $LaFeAsO_{1-x}H_x$, hydrogen acts as $H^-$, allowing it to overcome the poor solubility of $F^-$. The extended solubility limit led to the discovery of a two-dome-shaped superconducting phase diagram as a function of hydrogen concentration [9]. Furthermore, introduced hydrogen can behave not only as a dopant but also as a substantial impurity that enhances the disorder effect [10–13]. It is known that introducing nonmagnetic impurities is one effective method for determining whether a superconducting gap is fully gapped [14–17].

Recently, transition metal zirconide $CoZr_2$ has attracted attention because hydrogen incorporation drastically alters its ground state. While $CoZr_2$ is a superconductor with $T_c \sim 6$ K, hydrogenated $CoZr_2H_x$ with high hydrogen concentration ($x$ = 3.49 and 4.8) exhibits weak itinerant ferromagnetism instead of superconductivity [18,19]. Theoretical studies have proposed that the dramatic changes in the physical properties of $CoZr_2H_x$ are the result of a transition from Zr *4d* states dominance in the vicinity of the Fermi level ($E_F$) for $CoZr_2$ to Co *3d* at the ferromagnetic phase in $CoZr_2H_x$ [20]. In contrast, the effect of hydrogen incorporation on the superconducting state in the low-hydrogen-concentration regime, before the emergence of ferromagnetism, remains unclear.

In this study, we synthesized $CoZr_2H_x$ by annealing polycrystalline $CoZr_2$ in a hydrogen atmosphere with varied reaction times in order to control the hydrogen concentration. The resulting samples fell into two distinct concentration regimes: a low-concentration superconducting phase of $x$ = 0–0.054 with $I4/mcm$ (No. 140) structure and a high-concentration ferromagnetic phase of $x$ = 2.786 with $P4/ncc$ (No. 130) structure. Within the superconducting regime, we revealed a monotonic suppression of $T_c$ despite near-identical Debye temperatures and densities of states at the Fermi level. To elucidate the contrasting roles of hydrogen in the two phases, we performed Hard X-ray Photoelectron Spectroscopy (HAXPES) measurements on three representative compositions ($x$ = 0, 0.054, and 2.786). In the ferromagnetic phase ($x$ = 2.786), we found that the Zr *4d* metallic bands near $E_F$ are depleted via Zr–H hybridization, leaving Co *3d* states dominant at the

*Contact author: watanabe-yuto@ed.tmu.ac.jp
†Contact author: mizugu@tmu.ac.jp

Fermi level. In the superconducting phase ($x$ = 0 and 0.054), by contrast, the electronic structure remains largely unchanged upon hydrogenation. Because of no indication of ferromagnetic ordering and significantly altering the band structure, we confirmed that hydrogen behaves as a nonmagnetic impurity. The Abrikosov–Gor'kov pair-breaking theory described the $T_c$ suppression caused by hydrogenation, suggesting that the superconducting gap of $CoZr_2$ is not a fully isotropic $s$-wave but possesses an anisotropic or multi-gap structure as suggested in previous studies [21,22].

## II. METHODS

### A. Preparation of polycrystalline hydrogenated $CoZr_2H_x$ samples

Hydrogenated compounds of $CoZr_2H_x$ were synthesized by annealing polycrystalline $CoZr_2$ under a hydrogen gas atmosphere. The polycrystalline $CoZr_2$ was obtained by arc melting using Co wire (Nilaco, 99.995%) and Zr plates (Nilaco, 99.2%) as starting materials. The starting materials were placed on a water-cooled Cu hearth pocket, and a Ti ball was also placed in another pocket. The chamber was maintained under an argon atmosphere during the melting procedure. Before starting the melting procedure, the Ti ball was heated to absorb any residual oxygen left in the chamber. Then, starting materials were melted, forming a ball-shaped sample. The ball-shaped sample was turned over after each melting, and the melting was repeated several times to improve homogeneity. Following that, the obtained ball-shaped sample of $CoZr_2$ was ground to make it into fine powders. The fine powders wrapped by Mo foil (Nilaco, 99.95%) were annealed under hydrogen gas at 200°C under 0.1 MPa. To control the hydrogen concentration in the powders, the hydrogenation process was performed with reaction times of 10, 20, 30, 40, 50, and 60 minutes. Around 0.4 g of powder was used for each hydrogenation process. After the hydrogenation process, the powders containing a certain amount of hydrogen were finally sintered under high pressure in order to prevent hydrogen desorption during densification. The high-pressure sintering was performed by using a cubic-anvil type 180-ton press (C&T factory). Pelletized hydrogenated material with a 5 mm diameter was put in a high-pressure cell, which consists of a BN capsule, a carbon heater capsule, and a pyrophyllite cubic cell. The high-pressure sintering was performed at 300°C at 1.5 GPa for 20 minutes. The actual hydrogen concentrations were $x$ = 0.045(3), 0.054(1), 0.051(1), 0.040(1), 2.786(1), and 0.043(1) for 10, 20, 30, 40, 50, and 60 minutes of the hydrogenation process, respectively, confirmed through the thermal desorption spectroscopy (TDS) method. The TDS measurement results are shown in Fig. S1 in the supporting material [23].

### B. Characterization

Crystal structures of $CoZr_2H_x$ ($x$ = 0, 0.040, 0.043, 0.045, 0.051, 0.054, and 2.786) were confirmed through laboratory X-ray diffraction (XRD) by the $\theta$-$2\theta$ method with Cu-K$\alpha$ radiation on Miniflex600-C mounted on a D/teX Ultra detector. Collected XRD patterns were analyzed by the Rietveld method using RIETAN-FP [24]. Energy-dispersive X-ray spectroscopy was performed using a scanning electron microscope (TM4000plusIII) equipped with AZtecOne software, confirming that the actual chemical composition of Co and Zr is approximately 1:2.

Temperature dependence of magnetic susceptibility was measured using MPMS3 (Quantum Design) with a vibrating sample magnetometer mode. Temperature dependences of electrical resistivity and specific heat were measured using PPMS Dynacool (Quantum Design). The electrical resistivity measurements were performed by the conventional four-probe method with an alternating current of 2 mA. Ag paste (Dupont 4922N) and Au wire (Nilaco, 99.95%, φ = 0.025 mm) were used to build an electrical contact between the sample and the puck. The electrical resistivity was also measured under several magnetic fields between 0.05 and 9 T (0.05, 0.1, 0.15, 0.2, 0.3, 0.4, 0.5, 0.75, 1, 1.5, 2, 3, 6, and 9 T). The specific heat measurements were performed by the thermal relaxation method. Apiezon-N grease was used to get good thermal contact between the sample and the puck platform. The specific heat was measured under zero field and 9 T.

HAXPES measurements were performed at the Max-Planck-NSRRC HAXPES end station with an MB Scientific A-1 HE analyzer, at the Taiwan undulator beamline BL12XU of SPring-8 [25]. A photon energy of $h\nu$ = 6.5~keV with a resolution of around 250 meV was used. Measurements were performed at 80 K, and the pressure in the measuring chamber was of the order of $10^{-10}$ mbar. The samples were mounted on the sample holders coated with Ag paint and then fractured in-situ under an ultrahigh vacuum of the order of $10^{-9}$ mbar to expose a fresh surface for the measurements. Wide scans were performed to verify the lack of contaminants such as O, C, or Ag to ensure that the measurements were performed on clean, non-exposed parts of the sample. The additional bulk-sensitivity provided by the hard X-rays [26] allows us to obtain the signal from deeper inside the sample, minimizing the signal from the surface layers that might lose hydrogen during its exposure to ultra-high vacuum conditions.

*Contact author: watanabe-yuto@ed.tmu.ac.jp

†Contact author: mizugu@tmu.ac.jp

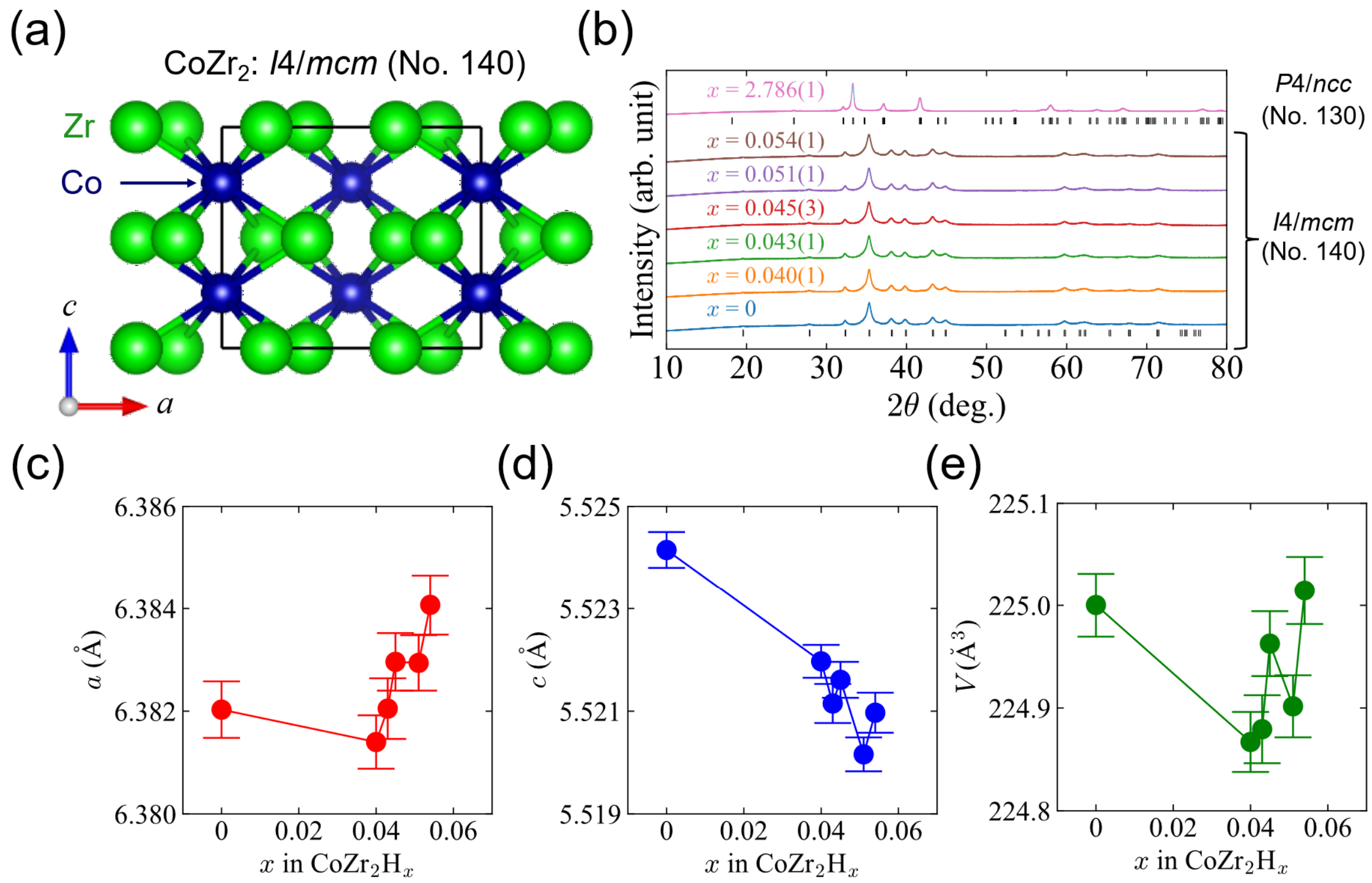

FIG 1. (a) Crystal structure image of $CoZr_2$ with a space group of *I*4/*mcm* (No. 140). The crystal structure is visualized using VESTA3 [31]. (b) Powder XRD patterns of hydrogenated $CoZr_2H_x$ ($x$ = 0, 0.040, 0.043, 0.045, 0.051, 0.054, and 2.786). (c–e) Hydrogen concentration dependences of lattice parameters of $a$ and $c$, and unit cell volume $V$.

Density Functional Theory (DFT) calculations of CoZr2 were performed using the full-potential local-orbital code FPLO 21 [27] with a k-mesh of 12×12×12 and scalar relativistic mode.

## III. RESULTS

### A. Crystal structure and lattice parameters

Pristine $CoZr_2$ has a tetragonal crystal structure with a space group *I*4/*mcm* (No. 140) as shown in Fig. 1(a). Co and Zr atoms occupy 4a (0,0,1/4) and 8h ($x$, $x$+0.5, 0) Wyckoff positions, respectively. The Zr layer and the Co layer are stacked along the $c$-axis. The Co atoms run along the $c$-axis, which forms a 1-dimensional Co chain. Figure 1(b) shows XRD patterns for $x$ = 0, 0.040, 0.043, 0.045, 0.051, 0.054, and 2.786. Compounds containing low amounts of hydrogen ($x$ = 0–0.054) also exhibit *I*4/*mcm* crystal structure. On the other hand, for $x$ = 2.786, the highest hydrogen concentration exhibits a *P*4/*ncc* (No. 130) structural XRD pattern. It is known that the crystal structure transforms to *P*4/*ncc* and exhibits itinerant ferromagnetism when $CoZr_2$ absorbs a relatively large amount of hydrogen [18–20]. In the present study, the hydrogen concentration remained as low as $x$ = 0.040 after 40 min of hydrogenation. However, extending the hydrogenation time by only 10 min resulted in the formation of $CoZr_2H_{2.786}$. Such an abrupt increase in hydrogen content suggests the presence of a gap in hydrogen concentration between the *I*4/*mcm* and *P*4/*ncc* phases, with the high-hydrogen *P*4/*ncc* phase becoming energetically favored above a certain hydrogen concentration. In fact, fully hydrogenated $CoZr_2H_5$ decomposes into $ZrH_2$ and $CoZrH_3$ in a dehydrogenation process [28]. This disproportionation behavior suggests that $CoZr_2H_x$ does not form a stable solid solution over a wide range of hydrogen concentrations. Instead, specific hydride phases are preferentially stabilized, which may account for the abrupt increase in hydrogen content observed in the present study. As for the lattice parameters, determined by Rietveld refinement, $a$ displays a slight increase with increasing hydrogen concentration (Fig.

*Contact author: watanabe-yuto@ed.tmu.ac.jp
†Contact author: mizugu@tmu.ac.jp

1c), while $c$ decreased with $x$ (Fig. 1d). Due to these opposite behaviors, the unit cell volume $V$ remains mostly constant with the hydrogenation (Fig. 1(e)). For $x$ = 0–0.054 with the *I*4/*mcm* structure, hydrogen was treated as occupying interstitial sites, and only the Co and Zr atoms were included in the Rietveld refinement, since laboratory X-ray sources are insensitive to the positions of light atoms such as hydrogen. On the other hand, for $x$ = 2.786 with the *P*4/*ncc* structure, we assumed hydrogen occupies 4b (3/4, 1/4, 0) and 16g (0.0329, 0.1645, 0.0761) Wyckoff positions, as determined by powder neutron diffraction [29]. Rietveld analysis results are shown in Fig. S2 in the supplemental material [30].

### B. Superconducting properties – Magnetic susceptibility and Resistivity

Next, we investigate the superconducting properties. Here, we are considering samples exhibiting superconductivity, namely those with $x$ = 0.040, 0.043, 0.045, 0.051, and 0.054. Figure 2(a) shows temperature dependences of magnetic susceptibility $4\pi\chi(T)$ measured under 1 mT. All samples exhibited a pronounced diamagnetic signal during zero-field cooling (ZFC). In contrast, in a field cooling (FC) protocol, the strength of the diamagnetic signal was suppressed. These behaviors are well known as typical type-II superconductors. Temperature dependences of $\chi(T)$ measured with the FC protocol under 0.1 T are shown in Fig. 2(b). The $\chi(T)$ retained Curie-Weiss-like paramagnetic behavior upon hydrogenation, with no indication of ferromagnetic ordering, suggesting that the introduced hydrogen does not induce a ferromagnetic moment. On the other hand, ferromagnetic temperature dependence of $\chi(T)$ was observed in $x$ = 2.786, as shown in Fig. S3(a) (see supplemental materials [32]). It is known that for $x$ = 3.49 and $x$ = 4.8, which have higher hydrogen concentrations, the Curie temperatures are 139 K [19] and 130 K [18], respectively. Figures 2(c) and 2(d) show temperature dependences of electrical resistivity $\rho(T)$ measured under zero field. Zero resistivity, in other words, superconducting transitions, were observed in all samples. According to the results of $4\pi\chi(T)$ and $\rho(T)$ measurements, $T_c$ slightly decreased by hydrogenation. Additionally, $\rho(T)$ data of $x$ = 2.786 measured under zero field are shown in Fig. S3(b) (see Supplemental Material [32]). The $\rho(T)$ of $x$ = 2.786 exhibited higher resistivity than that of $x$ = 0–0.054, consistent with previous studies [18,19]. For the normal-conducting states of $\rho(T)$, we fitted them using a parallel-resistor model expressed as follows:

$$\rho(T) = \left(\frac{1}{\rho_{\mathrm{BG}}(T)} + \frac{1}{\rho_{\mathrm{sat}}}\right)^{-1}. \qquad (1)$$

The $\rho_{\mathrm{BG}}(T)$ is a Bloch–Grüneisen resistivity model as follows: [33]

$$\rho_{\mathrm{BG}}(T) = \rho_{\mathrm{BG,0}} + \rho_{\mathrm{sd}}\left(\frac{T}{\Theta_{\mathrm{R}}}\right)^3 \int_0^{\Theta_{\mathrm{R}}/T} \frac{x^3}{(e^x-1)(1-e^{-x})}dx\,, \qquad (2)$$

where $\rho_{\mathrm{BG,0}}$ and $\rho_{\mathrm{sd}}$ are numerical constants, and $\Theta_{\mathrm{R}}$ is a characteristic temperature, typically the Debye temperature. Here, we assume a phonon-assisted *s*-*d* electron scattering model based on prior related research [34]. The $\rho_{\mathrm{sat}}$ is a term representing the saturating behavior of $\rho(T)$, which is realized when a mean free path becomes comparable to interatomic separations of the material, called the Ioffe–Regel condition [35]. The fits using Eqs. (1) and (2) reproduced the $\rho(T)$ data well. Table 1 shows the obtained fitting parameters and residual resistivity $\rho_0$. The $\rho_0$ can be calculated as follows:

$$\rho_0 = \frac{\rho_{\mathrm{BG,0}} \cdot \rho_{\mathrm{sat}}}{\rho_{\mathrm{BG,0}} + \rho_{\mathrm{sat}}}. \qquad (3)$$

The $\rho_0$ tends to increase with increasing hydrogen concentrations, which implies that electron scattering caused by hydrogen was enhanced.

We further measured $\rho(T)$ under several magnetic fields. Figure 2(e) shows the $\rho(T)$ data of $x$ = 0.054, which were measured under magnetic fields. $T_c$ gradually decreased with increasing magnetic fields, and $\rho(T)$ appears flat at 9 T down to 1.8 K. We obtained $T_c$ values at each magnetic field by using the 50% criterion for $\rho_0$, denoted as $T_c^{\rho(\mathrm{mid})}$. Figure 2(f) shows the temperature dependences of upper critical fields $\mu_0H_{c2}(T)$ derived from $T_c^{\rho(\mathrm{mid})}$. The $\mu_0H_{c2}(T)$ data were fitted by the commonly adopted phenomenological formula as follows: [36]

$$\mu_0H_{c2}(T) = \mu_0H_{c2}(0)\left\{\frac{1-\left(\frac{T}{T_c}\right)^2}{1+\left(\frac{T}{T_c}\right)^2}\right\}. \qquad (4)$$

$\mu_0H_{c2}(0)$ is the upper critical field at 0 K. From the fitting, we found that $CoZr_2$ exhibited the highest $\mu_0H_{c2}(0)$ value, and the hydrogenated compounds had values lower than that. $\mu_0H_{c2}(0)$ values are related to a length scale of superconductors, GL coherence length $\xi_{\mathrm{GL}}$. The $\xi_{\mathrm{GL}}$ can be expressed as follows:

$$\xi_{\mathrm{GL}} = \sqrt{\frac{\Phi_0}{2\pi\mu_0H_{c2}(0)}}, \qquad (5)$$

where $\Phi_0$ is a superconducting magnetic flux quantum. The calculated $\xi_{\mathrm{GL}}$ is shown in Table 2. The hydrogenated compounds exhibited slightly longer $\xi_{\mathrm{GL}}$ than $CoZr_2$. The $\xi_{\mathrm{GL}}$ contains not only the intrinsic superconducting coherence length but also the impurity scattering effect as shown below:

$$\frac{1}{\xi_{\mathrm{GL}}} = \frac{1}{\xi_0} + \frac{1}{l}, \qquad (6)$$

where $l$ is the mean free path. $\xi_0$ is the superconducting coherence length of pure materials derived from the

*Contact author: watanabe-yuto@ed.tmu.ac.jp
†Contact author: mizugu@tmu.ac.jp

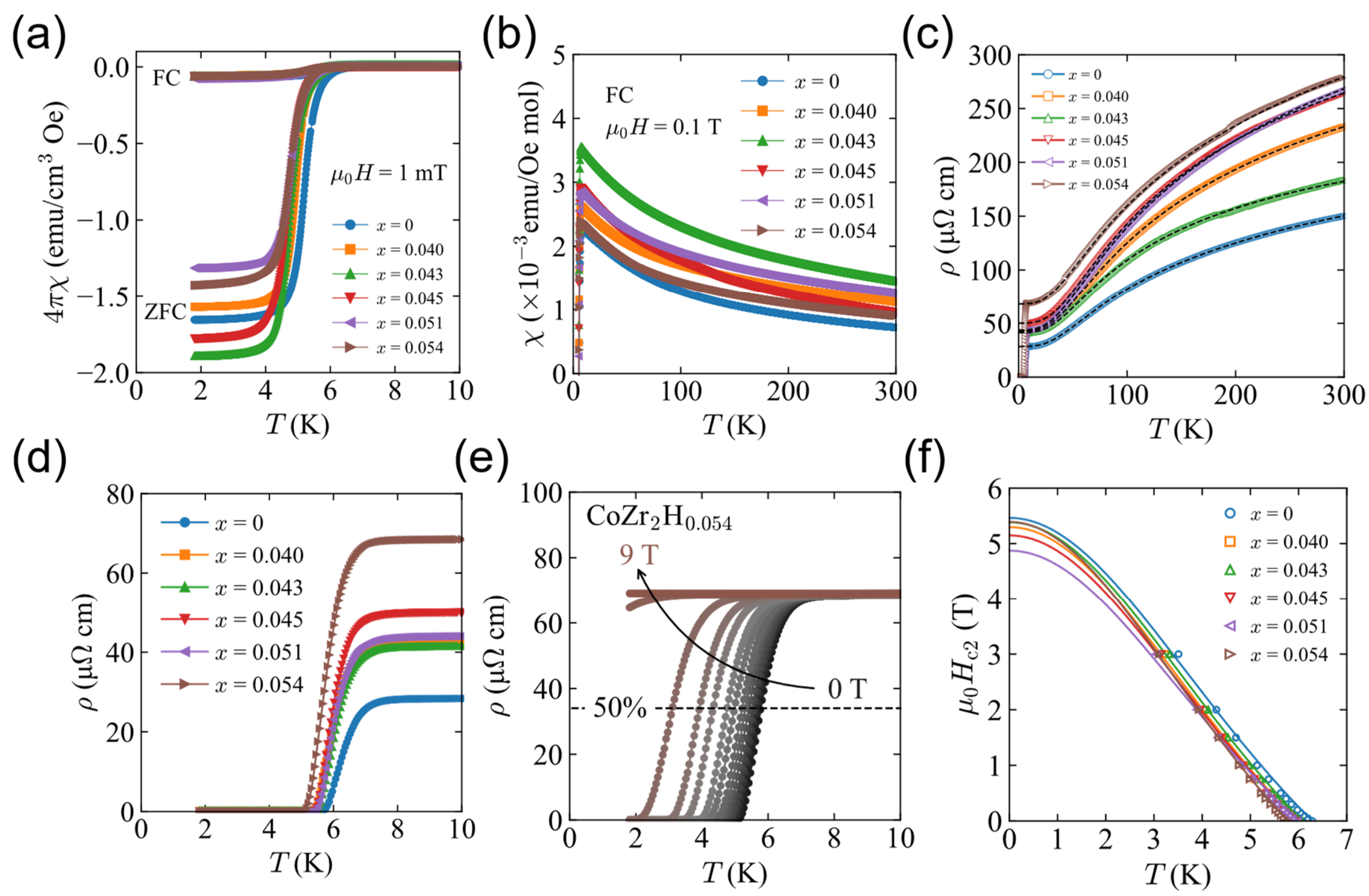

FIG. 2. (a) Temperature dependences of magnetic susceptibility with FC and ZFC protocols measured under 1 mT and (b) with FC protocol measured under 0.1 T. (c, d) Temperature dependences of electrical resistivity measured under zero field and near the $T_c$ region. The dashed lines represent fitting results by the parallel resistor model. (e) Temperature dependence of electrical resistivity for $x$ = 0.054 under several magnetic fields. The dashed line represents 50% criterion of residual resistivity. (f) Temperature dependences of upper critical field. The solid curves are fitting results with Eq. (4).

BCS theory, and it can be expressed with $\Delta(0)$ as follows: [37]

$$\xi_0 = \frac{\hbar v_F}{\pi \Delta(0)}. \tag{7}$$

$\hbar$ and $v_F$ are Dirac's constant and Fermi velocity, respectively. The impurity scattering effect generally influences the mean free path because of a relationship as $l \propto \rho_0^{-1}$. Since hydrogenated compounds have a higher $\rho_0$ than $CoZr_2$, as shown in Table 1, the $\xi_{GL}$ of them should be shorter than that of $CoZr_2$ if $\xi_0$ is maintained before and after hydrogenation. Therefore, the observed longer $\xi_{GL}$ on hydrogenated compounds suggests that $\xi_0$ becomes somewhat longer than $CoZr_2$. To elucidate pair-breaking mechanisms by magnetic field, we introduce two kinds of limits of upper critical fields: Pauli–Clogston limit $\mu_0 H_P$ and orbital limit $\mu_0 H_{orb}$. The $\mu_0 H_P$ can be expressed as follows by using the BCS theory: [37]

$$\mu_0 H_P = \frac{\Delta(0)}{\sqrt{2}\mu_B} = \frac{\pi k_B T_c e^{-\gamma_E}}{\sqrt{2}\mu_B} = 1.86 T_c \text{ [T]}, \tag{8}$$

where $k_B$, $\mu_B$, and $\gamma_E$ are the Boltzmann constant, Bohr magneton, and Euler constant, respectively. Additionally, an expression of the $\mu_0 H_{orb}$ in a dirty limit is shown as follows according to the Werthamer-Helfand-Hohenberg (WHH) theory: [38,39]

$$\mu_0 H_{orb} = -0.693 T_c \left.\frac{d\mu_0 H_{c2}(T)}{dT}\right|_{T=T_c}. \tag{9}$$

The ratio of these two limit values, known as the Maki parameter $\alpha_M = \sqrt{2}\mu_0 H_{orb}/\mu_0 H_P$, is used to determine which pair-breaking mechanism is dominant. The calculated parameters of $\mu_0 H_P$, $\mu_0 H_{orb}$, and $\alpha_M$ are also shown in Table 2. $\alpha_M$ shows values lower than $\sqrt{2}$ among all the samples, suggesting that the orbital pair-breaking effect is dominant.

*Contact author: watanabe-yuto@ed.tmu.ac.jp
†Contact author: mizugu@tmu.ac.jp

Table 1. Obtained fitting parameters for $\rho(T)$ derived from the parallel resistor model.

| sample | $\rho_{\text{BG},0}$ (μΩ cm) | $\rho_{\text{sd}}$ (μΩ cm) | $\Theta_{\text{R}}$(K) | $\rho_{\text{sat}}$ (μΩ cm) | $\rho_0$ (μΩ cm) |
|---|---|---|---|---|---|
| $CoZr_2$ | 31.82(5) | 554(2) | 247.5(6) | 259.1(3) | 28.34(4) |
| $CoZr_2H_{0.040}$ | 47.65(7) | 894(3) | 258.6(6) | 404.6(4) | 42.63(4) |
| $CoZr_2H_{0.043}$ | 48.7(2) | 933(7) | 266(1) | 271.0(4) | 41.26(4) |
| $CoZr_2H_{0.045}$ | 56.46(4) | 1121(2) | 273.6(4) | 445.8(3) | 50.12(4) |
| $CoZr_2H_{0.051}$ | 48.10(4) | 988(3) | 257.7(6) | 481.9(6) | 43.73(5) |
| $CoZr_2H_{0.054}$ | 78.97(4) | 918(3) | 235.0(5) | 490.3(5) | 68.02(4) |

Table 2. Superconducting parameters relating to the upper critical fields derived from $\rho(T)$ measurements under magnetic fields.

| sample | $T_c^{\rho(\text{mid})}$ (K) | $\mu_0H_{c2}(0)$ (T) | $\xi_{\text{GL}}$ (nm) | $\mu_0H_{\text{P}}$ (T) | $\mu_0H_{\text{orb}}$ (T) | $\alpha_{\text{M}}$ |
|---|---|---|---|---|---|---|
| $CoZr_2$ | 6.3(5) | 5.5(1) | 7.76(7) | 11.7(9) | 2.8(5) | 0.34(7) |
| $CoZr_2H_{0.040}$ | 6.0(5) | 5.30(8) | 7.88(6) | 11(1) | 2.8(5) | 0.36(8) |
| $CoZr_2H_{0.043}$ | 6.1(8) | 5.38(8) | 7.82(6) | 11(1) | 2.9(8) | 0.4(1) |
| $CoZr_2H_{0.045}$ | 6.0(3) | 5.15(7) | 8.00(6) | 11.1(6) | 2.9(3) | 0.36(5) |
| $CoZr_2H_{0.051}$ | 6.0(4) | 4.87(9) | 8.22(8) | 11.1(7) | 2.6(4) | 0.33(6) |
| $CoZr_2H_{0.054}$ | 5.8(3) | 5.39(4) | 7.81(3) | 10.7(6) | 3.2(3) | 0.42(5) |

### C. Superconducting properties – Specific heat

To investigate the superconducting properties in detail, we performed specific heat $C(T)$ measurements. Figure 3(a) shows temperature dependences of $C(T)/T$ measured under zero field and 9 T. All samples exhibited a superconducting jump, and the highest $T_c$ was observed on $CoZr_2$, which is consistent with $4\pi\chi(T)$ and $\rho(T)$ measurements. The 9 T data of $C(T)/T$ were fitted by using the following expression:

$$\frac{C(T)}{T} = \gamma + \beta T^2 + \delta T^4. \quad (10)$$

The coefficient of $\gamma$ is a Sommerfeld coefficient representing electrical contributions. The $\beta T^2+\delta T^4$ is attributed to phonon contributions. Electrical specific heat $C_{\text{el}}(T)$ under 0 T can be obtained by subtracting phonon contributions from the total specific heat, as shown in Fig. 3(b). We performed fitting for $C_{\text{el}}(T)$ data in the superconducting state by using a semiempirical approximation called the $\alpha$-model:

$$C_{\text{el}}(T) = A\exp[-\Delta(0)/k_{\text{B}}T], \quad (11)$$

where $A$ is a constant coefficient. Figure 3(c) shows the fitting result for $x$ = 0.054 by using the $\alpha$-model. $T_c$ values of specific heat, denoted as $T_c^C$ were determined by considering an entropy balance. From these fits, the Debye temperature $\Theta_{\text{D}}$ was derived by using $\beta$ as shown below:

$$\Theta_{\text{D}} = \left(\frac{12\pi^4 NR}{5\beta}\right)^{\frac{1}{3}}, \quad (12)$$

where $N$ and $R$ are the number of atoms in a formula unit and the gas constant, respectively. Additionally, the density of states at the Fermi level, considering spin degeneracy, DOS($E_{\text{F}}$), can be obtained through $\gamma$ as follows: [40]

$$\text{DOS}(E_{\text{F}}) = \frac{3\gamma}{\pi^2 k_{\text{B}}^2(1+\lambda_{\text{el-ph}})}. \quad (13)$$

To compensate for the increase in effective mass due to electron-phonon interactions, the electron-phonon coupling constant is introduced by using the McMillan relation: [40]

$$\lambda_{\text{el-ph}} = \frac{1.04+\mu^*\ln(\Theta_{\text{D}}/1.45T_{\text{c}})}{(1-0.62\mu^*)\ln(\Theta_{\text{D}}/1.45T_{\text{c}})-1.04}. \quad (14)$$

Here, we assume the Coulomb pseudopotential $\mu^*$ = 0.13, which is widely applied for compounds consisting of transition metals [40]. Figures 3(d)–3(f) show hydrogen concentration dependences of DOS($E_{\text{F}}$), $\Theta_{\text{D}}$, and $\Delta(0)$ derived from the above quantitative analyses. Other parameter values of the superconducting specific heat jump and electron-phonon coupling constant are listed in Table 3. The relatively small $\Delta C_{el}/\gamma T_c$ and $\lambda_{el\text{-}ph}$ values suggest weak-coupling superconductivity in $CoZr_2H_x$. As for $\Delta(0)$, the hydrogenated compounds exhibited smaller values than the dehydrogenated $CoZr_2$. A small $\Delta(0)$ results in longer $\xi_0$ as expressed in Eq. (7), the $\xi_{GL}$ of the hydrogenated compound becomes longer, even if the $l$ becomes shorter due to hydrogenation.

*Contact author: watanabe-yuto@ed.tmu.ac.jp
†Contact author: mizugu@tmu.ac.jp

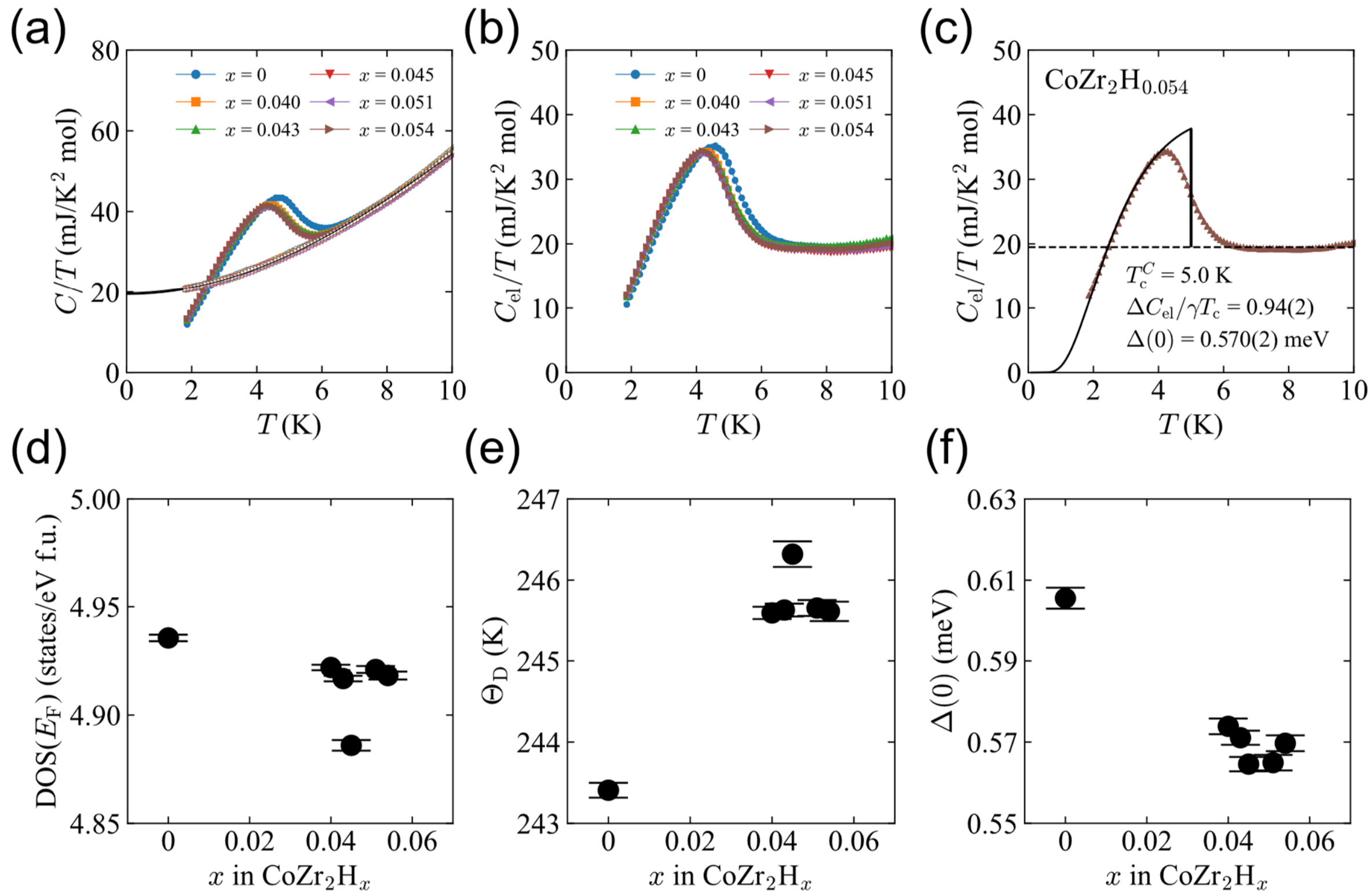

FIG. 3. (a) Temperature dependences of specific heat measured under zero field and 9 T. The zero-field data and 9 T data are shown as filled and open markers, respectively. The solid curves are the fit to the 9 T data obtained by using Eq. (10) (b) Temperature dependences of electrical specific heat. (c) Temperature dependence of electrical specific heat for $x = 0.054$ with a fit using an $\alpha$-model. (d–f) Hydrogen concentration dependences of density of states at the Fermi level, Debye temperature, and superconducting energy gap at 0 K.

Table 3. Superconducting parameters of $T_c^C$, $\Delta C_{el}/\gamma T_c$, and $\lambda_{el\text{-}ph}$.

| sample | $T_c^C$ (K) | $\Delta C_{el}/\gamma T_c$ | $\lambda_{el\text{-}ph}$ |
|---|---|---|---|
| $CoZr_2$ | 5.4 | 0.94(2) | 0.70 |
| $CoZr_2H_{0.040}$ | 5.1 | 0.94(2) | 0.69 |
| $CoZr_2H_{0.043}$ | 5.1 | 0.93(1) | 0.69 |
| $CoZr_2H_{0.045}$ | 5.0 | 0.94(1) | 0.68 |
| $CoZr_2H_{0.051}$ | 5.0 | 0.95(2) | 0.68 |
| $CoZr_2H_{0.054}$ | 5.0 | 0.94(2) | 0.68 |

## IV. Discussion

### A. Electronic structure

To clarify how hydrogen incorporation modifies the electronic structure, we performed HAXPES measurements on three representative compositions for the pristine $CoZr_2$ with *I*4/*mcm* phase, $CoZr_2H_{0.054}$ with lightly hydrogenated *I*4/*mcm* phase, and $CoZr_2H_{2.786}$ with heavily hydrogenated *P*4/*ncc* phase. Figure 4(a) shows the valence band spectra measurements for $x = 0$, 0.054, and 2.786. For $x = 0$, we observe that the spectra are well-described by the DFT calculations, with the peak crossing $E_F$ consisting mostly of Zr *4d* metallic bands, and the Co *3d* states deeper, centered in the 1–2 eV range. Meanwhile, for $x$ = 2.786, in the ferromagnetic phase, the Zr *4d* metallic states near the $E_F$ have almost entirely vanished, instead presenting now a significant additional spectral weight in the 6–8 eV region due to the Zr *4d* and H *1s* hybridization, in line with the theoretical models proposed in the literature [18,20]. For $x$ = 0.054, still within the superconducting phase, we observe that the spectrum is mostly identical to that of $x$ = 0. Here, instead of presenting energy shifts in the spectra that would be expected from a regular doping, the minor observed changes consisting of a very small loss in spectral weight at the Zr *4d*-derived feature at the $E_F$, together with a mild increase in the 6–8 eV region, where the Zr *4d*-H *1s* hybridized states are found for the ferromagnetic phase, suggesting some of the Zr ions bond with the H and start behaving more akin to what is observed in $x$ = 2.786 but not significantly affecting the rest otherwise. We can better understand this by examining the core-level spectra, shown in Fig. 4(b) and 4(c), which are often used to probe subtle chemical shifts and charge changes in the bands of intermetallic compounds [44,45]. The Co *2p* core level shows no noticeable shift all the way from $x$ = 0 to $x$ = 2.786, indicating that the inclusion of hydrogen does not have a relevant effect on the electronic occupation of the Co ions. Interestingly, at the $x$ = 2.786 ferromagnetic phase, the main peak becomes more asymmetric, a symptom suggesting an increased mobility of the Co *3d* electrons, i.e., more relevance at the $E_F$ and to its conductive properties. As for the Zr *3d* core level, from $x$ = 0 to 0.054, the main peak shows no noticeable change or shift, but a higher-energy, broader set of peaks appears at 183–185 eV (approx. 4 eV above the main peak). This can be understood as the contributions from higher valency Zr ions (and thus, less screened) resulting from the charge transfer of electrons from the Zr to the $H^+$ ions. The fact that, instead of shifting the main peak due to a uniform doping of the valence state of all Zr ions, there is now a small population of Zr ions with a significantly higher valence indicates thus that the extra charges of the H ions affect isolated Zr ions, as indeed proposed in the valence band discussions. Finally, at $x$ = 2.786, we observe not only that the peaks at 183–185 eV have increased, but also the main peaks have significantly shifted to deeper energies, showing thus that in the ferromagnetic phase the Zr ions increase in valence overall, shifting away from the $E_F$ as observed in the valence band spectra and also as hypothesized in the previous theoretical works.

*Contact author: watanabe-yuto@ed.tmu.ac.jp
†Contact author: mizugu@tmu.ac.jp

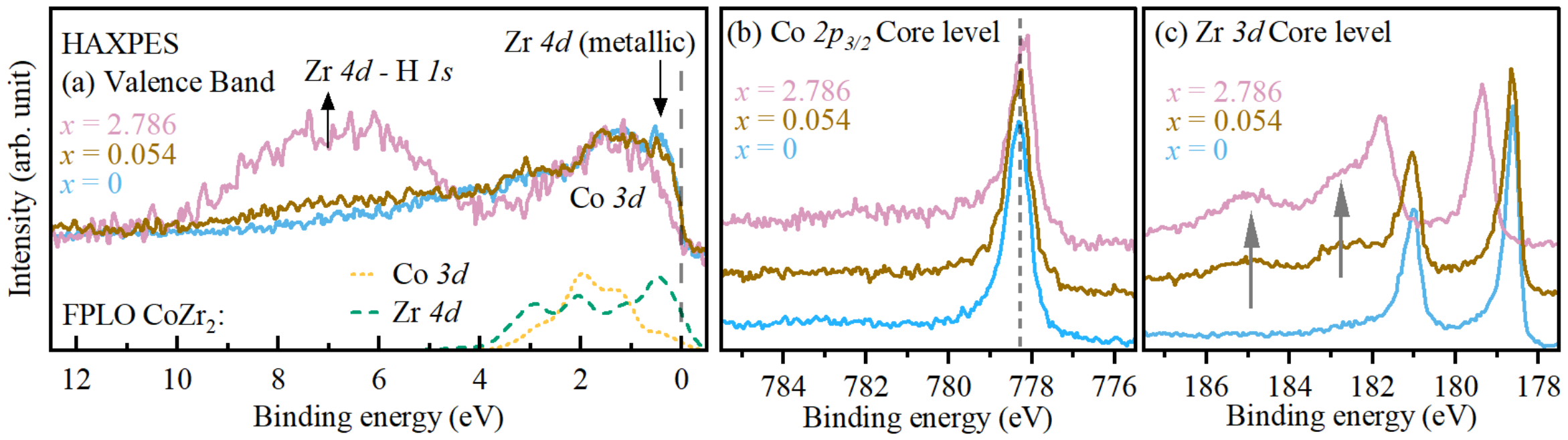

FIG 4. (a) HAXPES valence band spectra of $CoZr_2H_x$ ($x$ = 0, 0.054, and 2.786) together with the Co *3d* and Zr *4d* partial density of states (PDOS) calculated with FPLO. The PDOS have been multiplied by the Fermi function, broadened, and re-scaled according to the photoionization cross-sections to simulate the experimental conditions. HAXPES Co *2p* (b) and Zr *3d* (c) core level spectra of $CoZr_2H_x$.

We can thus understand the changes in electronic structure as follows: For low concentrations of hydrogen, its effects remain confined in their neighboring Zr ions without significantly altering the overall band structure and the character of the metallic bands crossing the Fermi level, resulting in an impurity-like effect in a still superconducting phase. However, with increasing $x$, the Zr metallic bands are gradually depleted, leaving, ultimately, mainly the Co *3d* states at the Fermi level to become dominant in the conducting bands. This results in different orbital degrees of freedom allowing for different hopping and band formations, which might energetically favor a distinct ion arrangement, i.e., crystal structure phase.

### B. Suppression of $T_c$

Finally, we discuss how introduced hydrogen can suppress the superconducting transition temperatures for the $CoZr_2H_x$ system. Figure 5(a) shows the dependence of $T_c$ on hydrogen concentration, determined from $\rho(T)$ ($T_c^{\rho(\mathrm{mid})}$), $4\pi\chi(T)$ ($T_c^{\chi}$), and $C(T)$ ($T_c^{C}$) measurements. As discussed above, $T_c$ values slightly decrease with increasing hydrogen concentration, even near-identical $\Theta_D$ and DOS($E_F$), as we can see in Figs 3(d) and 3(e). According to Anderson's theorem for superconductivity [41], $T_c$ values of conventional superconductors that possess a spherical superconducting $s$-wave gap structure are not influenced by nonmagnetic impurities. However, Anderson's theorem cannot be applied to unconventional superconductors with a nodal or anisotropic superconducting gap. In fact, previous studies have suggested that the superconducting gap structure of $CoZr_2$ seems to be anisotropic $s$-wave-like or a multi-gap structure [21, 22]. Suppression of $T_c$ caused by nonmagnetic impurities can be described by the Abrikosov–Gor'kov (AG) pair-breaking theory [42].

*Contact author: watanabe-yuto@ed.tmu.ac.jp
†Contact author: mizugu@tmu.ac.jp

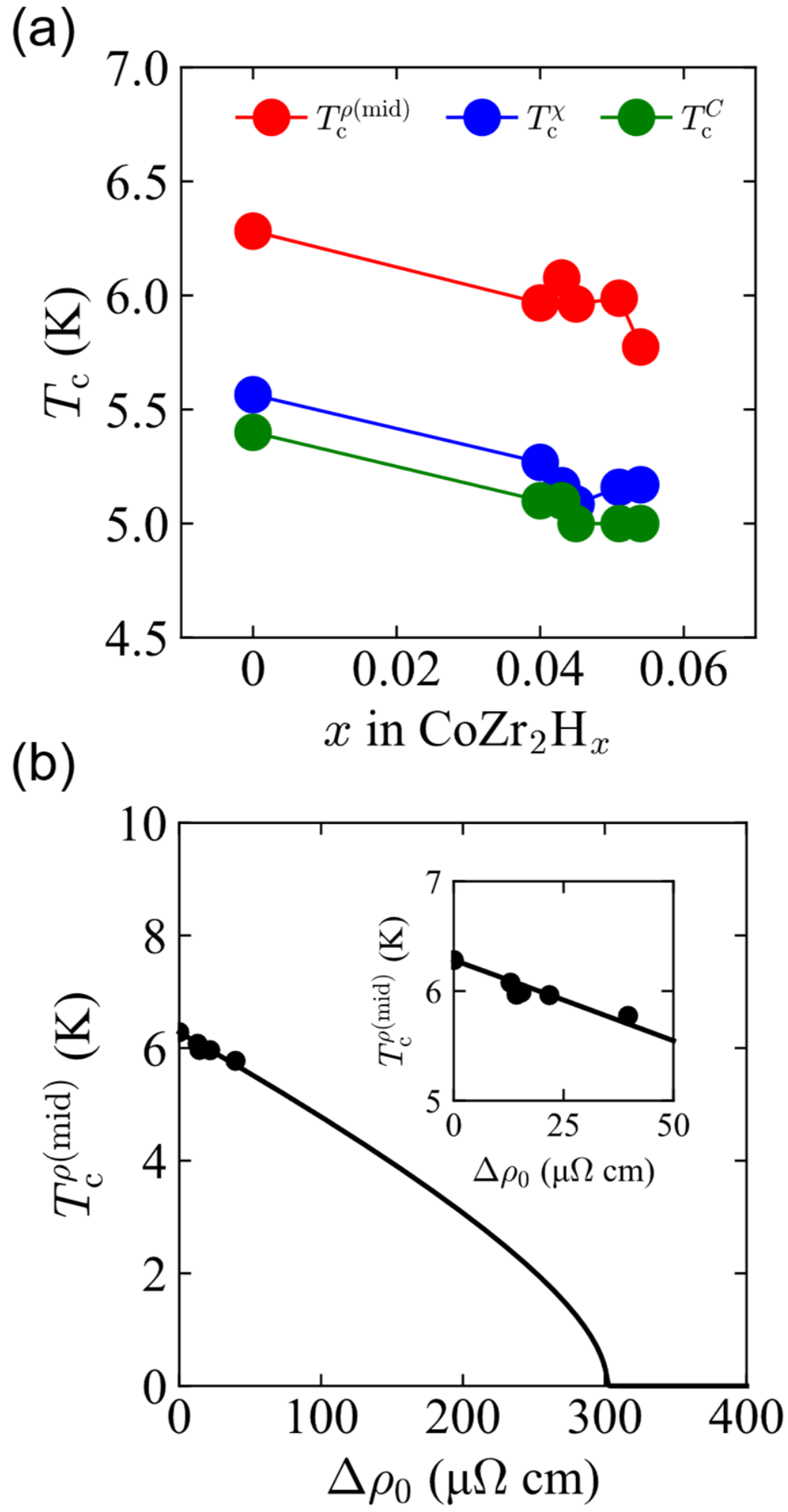

FIG 5. (a) Hydrogen concentration dependence of superconducting transition temperatures determined by resistivity, magnetic susceptibility, and specific heat measurements. (b) Residual resistivity dependence of superconducting transition temperature. The solid curve represents a fitting result by the AG formula.

*Contact author: watanabe-yuto@ed.tmu.ac.jp
†Contact author: mizugu@tmu.ac.jp

According to the AG theory, $T_c$ values obey the following expression:

$$\ln\left(\frac{T_{c0}}{T_c}\right) = \Psi\left(\frac{1}{2} + \frac{\alpha T_{c0}}{2\pi T_c}\right) - \Psi\left(\frac{1}{2}\right). \quad (15)$$

The $\Psi$ is the digamma function. $T_{c0}$ is the superconducting transition temperature in the absence of impurity scattering. The pair-breaking parameter $\alpha$ is expressed as $\alpha = \hbar/2\tau k_B T_{c0}$, where $\tau$ is the mean free time for impurity scattering. Therefore, $\alpha = 0$ leads to $T_c = T_{c0}$. To conduct quantitative analyses for the $CoZr_2H_x$ system, we fit $T_c^{\rho(\mathrm{mid})}$ data to the following generalized AG formula: [43]

$$\ln\left(\frac{T_{c0}}{T_c}\right) = \Psi\left(\frac{1}{2} + C_{AG}\frac{\Delta\rho_0}{T_c}\right) - \Psi\left(\frac{1}{2}\right). \quad (16)$$

The coefficient $C_{AG}$ is a scaling parameter. $\Delta\rho_0$ is determined as $\Delta\rho_0 = \rho_0 - \rho_0(CoZr_2)$ to consider only the effect of impurity scattering caused by the introduced hydrogen. The fitting result using the AG formula is shown in Fig. 5(b). The fitting parameter $C_{AG}$ was estimated as 0.0029(3) K/μΩ cm. The suppression of $T_c$ by the introduced hydrogen depends linearly on $\Delta\rho_0$ within a small pair-breaking factor, and the behavior is described by the Abrikosov–Gor'kov formula. The AG-like suppression of $T_c$ suggests that the superconducting gap deviates from fully opened $s$-wave symmetry.

## V. CONCLUSION

In summary, we synthesized $CoZr_2H_x$ by systematically varying the hydrogenation annealing time and found that the resulting samples segregate into two distinct concentration regimes: a low-concentration superconducting phase ($x$ = 0–0.054, *I*4/*mcm* structure) and a high-concentration ferromagnetic phase ($x$ = 2.786, *P*4/*ncc* structure). Within the superconducting regime, the hydrogenated compounds exhibit a monotonic suppression of $T_c$ despite near-identical Debye temperatures and densities of states at the Fermi level. The magnetic susceptibility of the low-concentration series shows no sign of ferromagnetic ordering.

To elucidate how hydrogenation influences the electronic structure in each regime, we performed HAXPES measurements on three representative compositions ($x$ = 0, 0.054, and 2.786). In the ferromagnetic phase ($x$ = 2.786), the Zr *4d* metallic bands near $E_F$ are substantially depleted via Zr–H hybridization, and Co *3d* states become dominant at the Fermi level. In the superconducting phase ($x$ = 0 and 0.054), by contrast, the electronic structure remains largely unchanged upon hydrogenation, confirming that hydrogen offers an impurity-like effect.

Given the nonmagnetic-impurity nature of hydrogen, the observed $T_c$ suppression in the superconducting regime was described by the Abrikosov–Gor'kov pair-breaking theory. Since Anderson's theorem dictates that nonmagnetic impurities leave $T_c$ unchanged in a fully-gapped *s*-wave superconductor, the $T_c$ reduction caused by hydrogenation indicates that the superconducting gap of $CoZr_2$ is not of the conventional fully-gapped *s*-wave type but rather of an unconventional type.

## ACKNOWLEDGMENTS

We acknowledge the support for the HAXPES measurements from the Max Planck-POSTECH-Hsinchu Center for Complex Phase Materials. D. T. acknowledges the support and assistance from M. Yoshimura for the HAXPES measurements. This work was supported by ERATO “Magnetic Thermal Management Materials” (No. JPMJER2201) from the Japan Science and Technology Agency, and Grant-in-Aid for Scientific Research (KAKENHI) (No. JP26K01198, 23KK0088, and JP25KJ1992) from Japan Society for the Promotion of Science. T. K. was supported by a project of Kanagawa Institute of Industrial Science and Technology (KISTEC) and Special Award for Science Tokyo Advanced Researchers (STAR) funded by Institute of Science Tokyo. The authors used Claude (Anthropic) to assist with language editing and refinement of the manuscript text.

[1] H. M. Syed, C. J. Webb, and E. MacA. Gray, Hydrogen-modified superconductors: A review, Prog. Solid State Chem. **44**, 20 (2016).

[2] A. Pundt and R. Kirchheim, HYDROGEN IN METALS: Microstructural Aspects, Annu. Rev. Mater. Res. **36**, 555 (2006).

[3] E. S. Kooij, A. T. M. Van Gogh, D. G. Nagengast, N. J. Koeman, and R. Griessen, Hysteresis and the single-phase metal-insulator transition in switchable $YH_x$ films, Phys. Rev. B **62**, 10088 (2000).

[4] T. Kawae, Y. Inagaki, S. Wen, S. Hirota, D. Itou, and T. Kimura, Superconductivity in Palladium Hydride Systems, J. Phys. Soc. Jpn. **89**, 051004 (2020).

[5] T. Skoskiewicz, Superconductivity in the palladium-hydrogen and palladium-nickel-hydrogen systems, Phys. Status Solidi a, **11**, K123 (1972).

[6] R. W. Standley, M. Steinback, and C. B. Satterthwaite, Superconductivity in $PdH_x(D_x)$ from 0.2 K to 4 K, Solid State Commun. **31**, 801 (1979).

[7] K. Miyazawa et al., Possible hydrogen doping and enhancement of $T_c$ (=35 K) in a LaFeAsO-based superconductor, Appl. Phys. Lett. **96**, 072514 (2010).

[8] S. Matsuishi, T. Hanna, Y. Muraba, S. W. Kim, J. E. Kim, M. Takata, S. Shamoto, R. I. Smith, and H. Hosono, Structural analysis and superconductivity of $CeFeAsO_{1-x}H_x$, Phys. Rev. B **85**, 014514 (2012).

[9] S. Iimura, S. Matsuishi, H. Sato, T. Hanna, Y. Muraba, S. W. Kim, J. E. Kim, M. Takata, and H. Hosono, Two-dome structure in electron-doped iron arsenide superconductors, Nat. Commun. **3**, 943 (2012).

[10] B. I. Belevtsev and V. I. Odnokozov, Influence of hydrogen impurities on the electrical and superconducting properties of disordered lead films, Sov. J. Low Temp. Phys. **11**, 249 (1985).

[11] C. Zhang, C.-X. Zhang, Y. Cui, R. Cao, S.-H. Wei, and H.-X. Deng, Role of hydrogen impurities in the high-$T_c$ superconductivity of infinite-layer nickelates, Phys. Rev. B **112**, 054512 (2025).

[12] K. Miyakawa, H. Takata, T. Yamaguchi, Y. Inagaki, K. Makise, and T. Kawae, Hydrogen-impurity-induced conductance peaks in constriction type Josephson junctions, Appl. Phys. Express **15**, 013002 (2021).

[13] C. Nölscher and G. Saemann-Ischenko, Superconductivity and crystal and electronic structures in hydrogenated and disordered $Nb_3Ge$ and $Nb_3Sn$ layers with A15 structure, Phys. Rev. B **32**, 1519 (1985).

[14] H. Alloul, P. Mendels, H. Casalta, J. F. Marucco, and J. Arabski, Correlations between magnetic and superconducting properties of Zn-substituted $YBa_2Cu_3O_{6+x}$, Phys. Rev. Lett. **67**, (1991).

[15] A. P. Mackenzie, R. K. W. Haselwimmer, A. W. Tyler, G. G. Lonzarich, Y. Mori, S. Nishizaki, and Y. Maeno, Extremely Strong Dependence of Superconductivity on Disorder in $Sr_2RuO_4$, Phys. Rev. Lett. **80**, (1998).

[16] Y. Fukuzumi, K. Mizuhashi, K. Takenaka, and S. Uchida, Universal Superconductor-Insulator Transition and $T_c$ Depression in Zn-Substituted High-$T_c$ Cuprates in the Underdoped Regime, Phys. Rev. Lett. **76**, 684 (1996).

[17] K. Karpińska, M. Z. Cieplak, S. Guha, A. Malinowski, T. Skośkiewicz, W. Plesiewicz, M. Berkowski, B. Boyce, T. R. Lemberger, and P. Lindenfeld, Metallic Nonsuperconducting Phase and *D*-Wave Superconductivity in Zn-Substituted $La_{1.85}Sr_{0.15}CuO_4$, Phys. Rev. Lett. **84**, 155 (2000).

[18] H. Mizoguchi, S. Matsuishi, S.-W. Park, and H. Hosono, Hydrogen-Insertion-Induced Itinerant

*Contact author: watanabe-yuto@ed.tmu.ac.jp

†Contact author: mizugu@tmu.ac.jp

Ferromagnetism in $Zr_2CoH_{4.8}$ with Co Chains, J. Phys. Chem. C **123**, 14964 (2019).
[19] Y. Watanabe, K. Suzuki, T. Katase, A. Miura, A. Yamashita, and Y. Mizuguchi, Uniaxial Negative Thermal Expansion in a Weak-Itinerant-Ferromagnetic Phase of $CoZr_2H_{3.49}$, J. Am. Chem. Soc. **148**, 7895 (2026).
[20] S. F. Matar, Drastic changes of electronic, magnetic, mechanical and bonding properties in $Zr_2Co$ by hydrogenation, Intermetallics **36**, 25 (2013).
[21] A. Teruya *et al.*, Superconducting and Fermi Surface Properties of Single Crystal $Zr_2Co$, J. Phys. Soc. Jpn. **85**, 034706 (2016).
[22] T. Matsuo, H. Sugita, S. Wada, and M. Ishikawa, Superconductivity in an itinerant nearly antiferromagnetic $Zr_2(Co_{1-x}Ni_x)$, Physica B: Condensed Matter **284–288**, 475 (2000).
[23] Y. Watanabe *et al.*, See Supplemental Material at [URL] for the TDS measurement results.
[24] F. Izumi and K. Momma, Three-Dimensional Visualization in Powder Diffraction, Solid State Phenom. **130**, 15 (2007).
[25] D. Takegami *et al.*, Valence band hard x-ray photoelectron spectroscopy on 3*d* transition-metal oxides containing rare-earth elements, Phys. Rev. B **99**, 165101 (2019).
[26] J. C. Woicik (ed.), Hard X-ray Photoelectron Spectroscopy (HAXPES), Springer Series in Surface Sciences (Springer International Publishing, Cham, 2016).
[27] K. Koepernik and H. Eschrig, Full-potential nonorthogonal local-orbital minimum-basis band-structure scheme, Phys. Rev. B **59**, 1743 (1999).
[28] Y. Liu *et al.*, Enhancing disproportionation resistance of $Zr_2Co$-based alloys by regulating the binding energy of H atom, Renew. Energy **233**, 121153 (2024).
[29] A. B. Riabov, V. A. Yartys, and H. Fjellva, Neutron diffraction studies of Zr-containing intermetallic hydrides with ordered hydrogen sublattice. V. Orthorhombic $Zr_3CoD_{6.9}$ with filled $Re_3B$-type structure, J. Alloys Compd. **296**, 312–316(2000).
[30] Y. Watanabe *et al.*, See Supplemental Material at [URL] for the Rietveld analysis results.
[31] K. Momma and F. Izumi, VESTA 3 for three-dimensional visualization of crystal, volumetric and morphology data, J. Appl. Crystallogr. **44**, 1272 (2011).
[32] Y. Watanabe *et al.*, See Supplemental Material at [URL] for the temperature dependences of $\chi(T)$ and $\rho(T)$ with $x = 2.786$.

*Contact author: watanabe-yuto@ed.tmu.ac.jp
†Contact author: mizugu@tmu.ac.jp

[33] A. H. Wilson, The electrical conductivity of the transition metals, Proc. R. Soc. Lond. A **167**, 580 (1938).
[34] Y. Watanabe *et al.*, Pressure-Induced Volumetric Negative Thermal Expansion in $CoZr_2$ Superconductor, Adv. Electron. Mater. **10**, 2300896 (2024).
[35] O. Gunnarsson, M. Calandra, and J. E. Han, Colloquium : Saturation of electrical resistivity, Rev. Mod. Phys. **75**, 1085 (2003).
[36] I. Kubo, Y. Watanabe, S. Kawashima, T. Miyaji, Y. Mizuguchi, T. Nishizaki, and J. Kitagawa, Enhanced $T_c$ in eutectic high-entropy alloy superconductors Hf-Nb-Sc-Ti-Zr, J. Alloys Compd. **1044**, 184531 (2025).
[37] M. Tinkham, Introduction to Superconductivity (Courier Corporation, North Chelmsford, 1996).
[38] N. R. Werthamer, E. Helfand, and P. C. Hohenberg, Temperature and Purity Dependence of the Superconducting Critical Field, $H_{c2}$. III. Electron Spin and Spin-Orbit Effects, Phys. Rev. **147**, 295 (1966).
[39] L. J. Neuringer and Y. Shapira, Effect of Spin-Orbit Scattering on the Upper Critical Field of High-Field Superconductors, Phys. Rev. Lett. **17**, 81 (1966).
[40] W. L. McMillan, Transition Temperature of Strong-Coupled Superconductors, Phys. Rev. **167**, 331 (1968).
[41] P. W. Anderson, Theory of dirty superconductors, J. Phys. Chem. Solids **11**, 26 (1959).
[42] A. A. Abrikosov and L. P. Gor'kov, Contribution to the theory of superconducting alloys with paramagnetic impurities. Sov. Phys. JETP **12**, 1243 (1961).
[43] A. Ranna *et al.*, Disorder-Induced Suppression of Superconductivity in Infinite-Layer Nickelates, Phys. Rev. Lett. **135**, 126501 (2025).
[44] I. Antonyshyn, O. Sichevych, U. Burkhardt, A. M. B. Jiménez, A. Melendez-Sans, Y.-F. Liao, K.-D. Tsuei, D. Kasinathan, D. Takegami, and A. Ormeci, Al–Pt intermetallic compounds: HAXPES study, Phys. Chem. Chem. Phys. **25**, 31137 (2023).
[45] M. Asai, H. Miyazaki, K. Watanabe, A. Yasui, Y. Takagi, and Y. Nishino, Hard X-Ray Photoemission Study of Heusler-Type $Fe_{2-x}Re_xVAl$ Thermoelectric Compounds, Phys. Status Solidi B **259**, 2100567 (2022).

Supplemental Materials

# Distinct Roles of Hydrogen in Superconducting and Ferromagnetic Phases of $CoZr_2H_x$

Yuto Watanabe,[1]* Takayoshi Katase,[2,3] Daisuke Takegami,[1,4] Yoshikazu Mizuguchi[1]†

*[1]Department of Physics, Tokyo Metropolitan University, 1-1 Minami-Osawa, Hachioji, Tokyo 192-0397, Japan*
*[2] Materials and Structures Laboratory, Institute of Integrated Research, Institute of Science Tokyo, 4259 Nagatsuta, Midori-ku, Yokohama, Kanagawa 226-8501, Japan*
*[3] MDX Research Center for Element Strategy, Institute of Integrated Research, Institute of Science Tokyo, 4259 Nagatsuta, Midori-ku, Yokohama, Kanagawa 226-8501, Japan*
*[4]Max Planck Institute for Chemical Physics of Solids, Nöthnitzer Straße 40, 01187 Dresden, Germany*

Actual concentration of hydrogen in $CoZr_2H_x$ was confirmed through thermal desorption spectrometry (TDS) measurements. Figure S1 shows the TDS measurement results for $CoZr_2H_x$ prepared by annealing at 200°C in a hydrogen gas atmosphere with 0.1 MPa. The hydrogen concentration was controlled by changing the hydrogenation annealing time. Figures S1(a)–S1(f) correspond to compounds which annealed for 40, 60, 10, 30, 20, and 50 minutes, respectively. We heated powder samples to 1000°C, and the hydrogen concentrations were estimated by a mass spectrometer, considering $m/z$ = 1 ($H^+$), $m/z$ = 2 ($H_2$), and $m/z$ = 18 ($H_2O$).

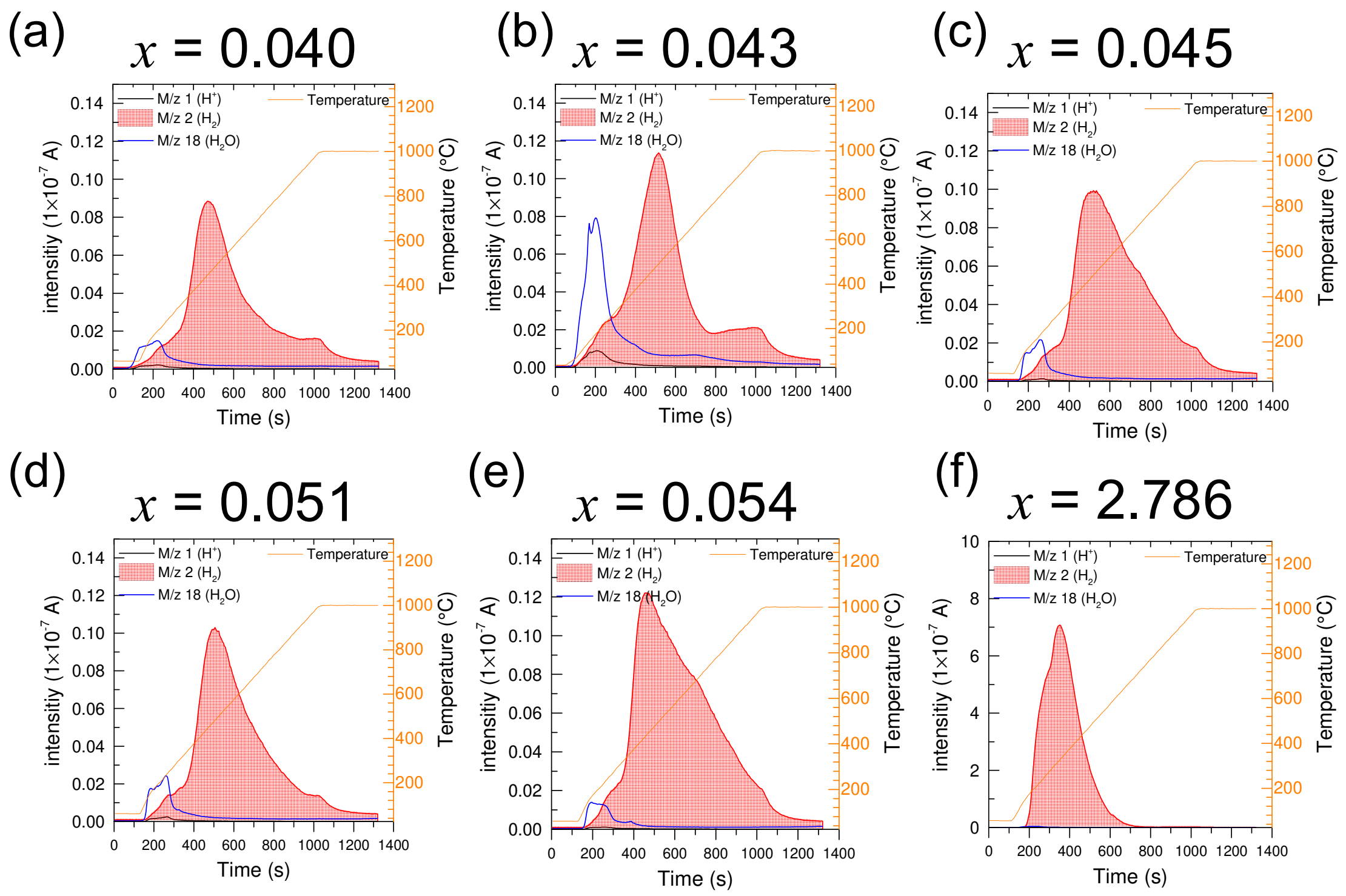

Figure S1. TDS measurement results of $CoZr_2H_x$ for (a) $x$ = 0.040, (b) $x$ = 0.043, (c) $x$ = 0.045, (d) $x$ = 0.051, (e) $x$ = 0.054, and (f) $x$ = 2.786.

*Contact author: watanabe-yuto@ed.tmu.ac.jp

†Contact author: mizugu@tmu.ac.jp

Figure S2 shows Rietveld refinement results for $CoZr_2H_x$ compounds. The retrieved analyses were performed by RIETAN-FP [24]. For $x$ = 0–0.054 exhibiting superconductivity, the dominant phase is confirmed as $CoZr_2H_x$ with *I*4/*mcm* (No. 140) structure, containing around 83–85% of mass fraction. Other phases were assumed to be $CoZr_3$ with *Cmcm* (No. 63) structure and CoZr with $Pm\bar{3}m$ (No. 221) structure. For $x$ = 2.786 in the absence of superconductivity, the crystal structure was determined to be *P*4/*ncc* (No. 130), and other impurity phases were almost negligible.

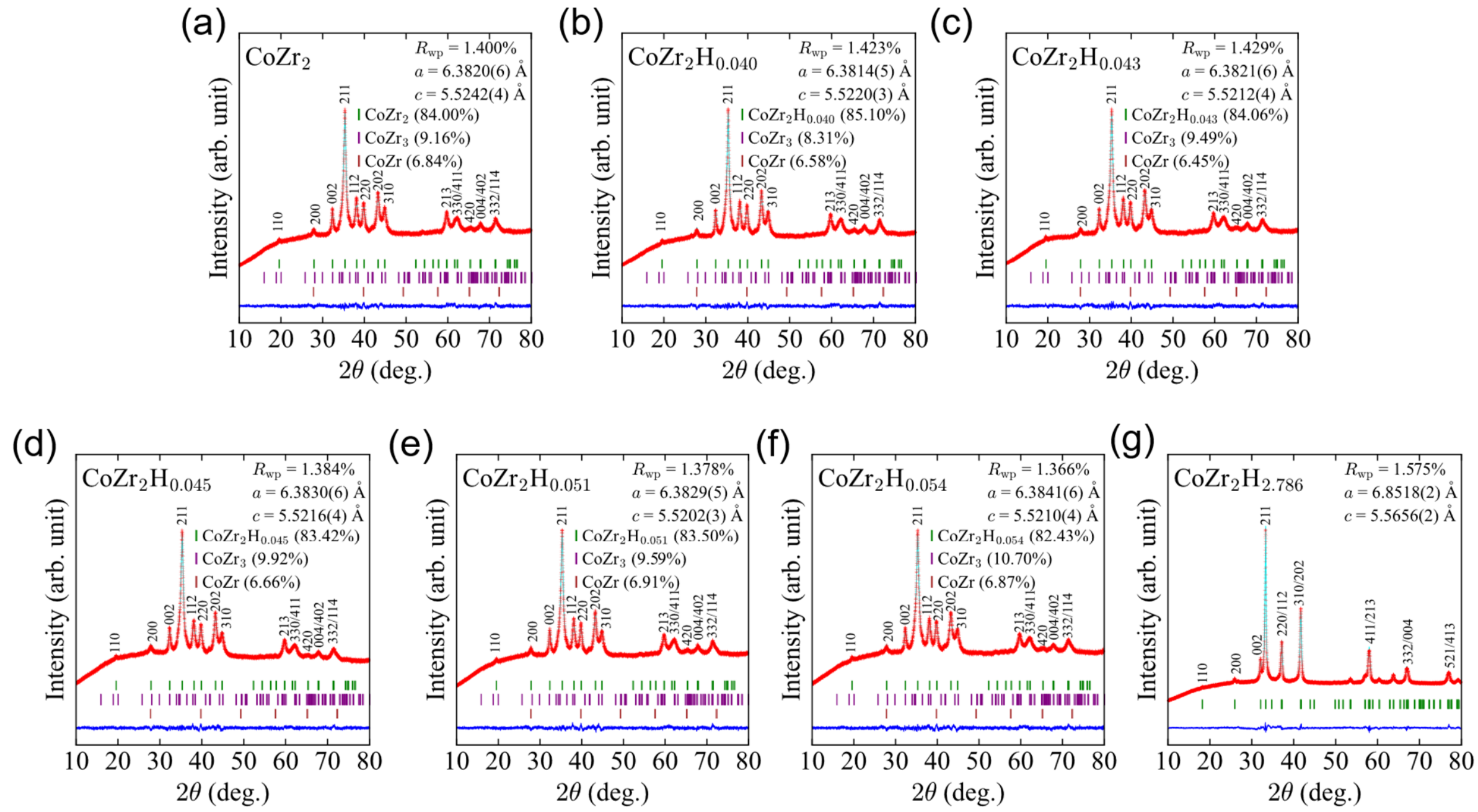

Figure S2. Rietveld refinement results of $CoZr_2H_x$ for (a) $x$ = 0, (b) $x$ = 0.040, (c) $x$ = 0.043, (d) $x$ = 0.045, (e) $x$ = 0.051, (f) $x$ = 0.054, and (g) $x$ = 2.786. We applied *I*4/*mcm* (No. 140) and *P*4/*ncc* (No. 130) space groups for $x$ = 0–0.054, and $x$ = 2.786, respectively.

*Contact author: watanabe-yuto@ed.tmu.ac.jp

†Contact author: mizugu@tmu.ac.jp

To compare the temperature dependences of magnetic susceptibility $\chi(T)$ and electrical resistivity $\rho(T)$ between $x$ = 0–0.054 exhibiting superconductivity and $x$ = 2.786 in the absence of superconductivity, we show $\chi(T)$ and $\rho(T)$ behaviors with $x$ = 2.786 in Figs. S3(a) and S3(b). The value of $\chi(T)$ for $x$ = 2.786 was higher than that of others, agreeing with previous studies with $x$ = 3.49 (Curie temperature is 139 K) [19] and 4.8 (Curie temperature is 130 K) [18].

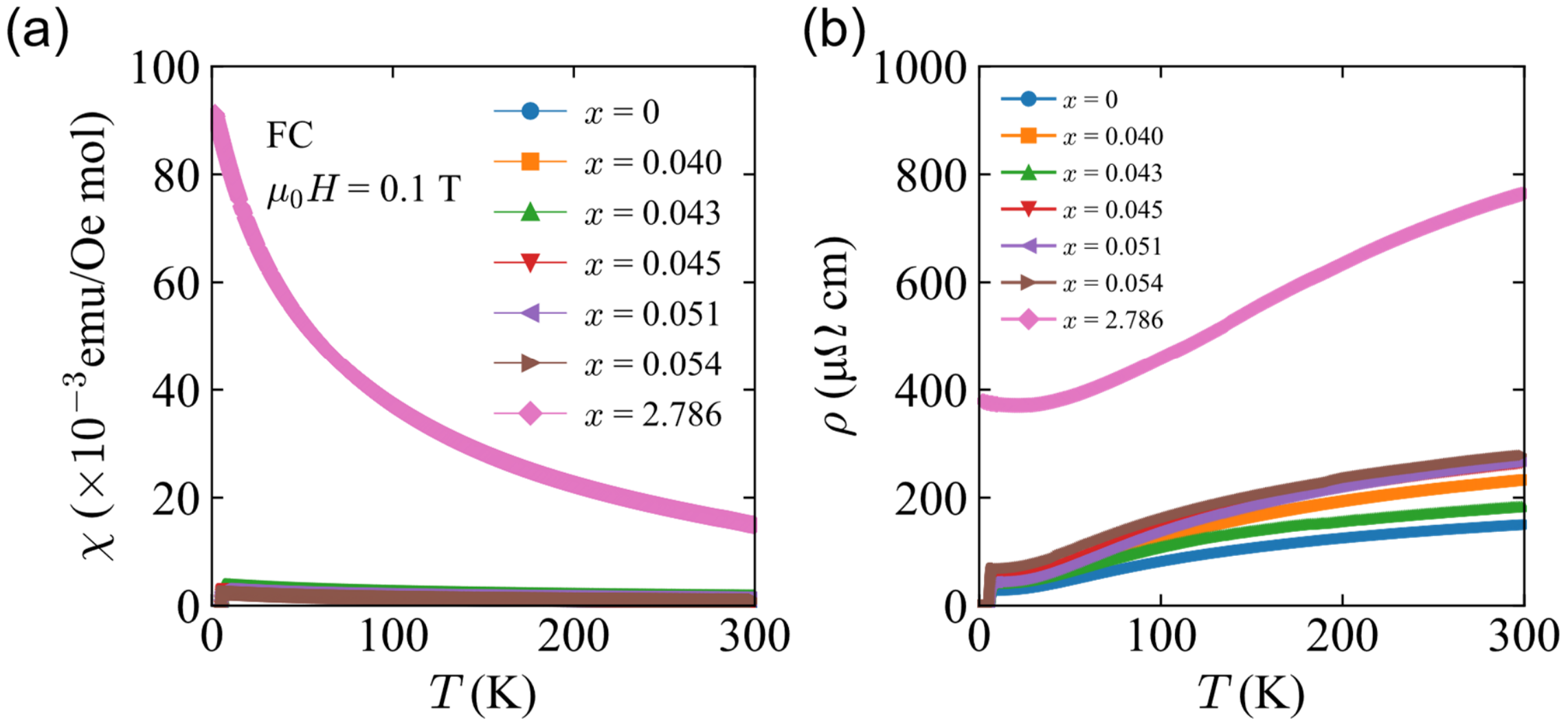

Figure S3. (a) Temperature dependences of magnetic susceptibility with $x$ = 2.786 exhibiting ferromagnetism. (b) Temperature dependences of electrical resistivity $\rho(T)$ with $x$ = 2.786 without superconductivity.

References

[18] H. Mizoguchi, S. Matsuishi, S.-W. Park, and H. Hosono, Hydrogen-Insertion-Induced Itinerant Ferromagnetism in $Zr_2CoH_{4.8}$ with Co Chains, J. Phys. Chem. C 123, 14964 (2019).

[19] Y. Watanabe, K. Suzuki, T. Katase, A. Miura, A. Yamashita, and Y. Mizuguchi, Uniaxial Negative Thermal Expansion in a Weak-Itinerant-Ferromagnetic Phase of $CoZr_2H_{3.49}$, J. Am. Chem. Soc. 148, 7895 (2026)

[24] F. Izumi and K. Momma, Three-Dimensional Visualization in Powder Diffraction, Solid State Phenom. 130, 15 (2007).

*Contact author: watanabe-yuto@ed.tmu.ac.jp

†Contact author: mizugu@tmu.ac.jp